# Quest-V: A Virtualized Multikernel for Safety-Critical Real-Time Systems

Richard West, Ye Li and Eric Missimer

*Computer Science Department*
*Boston University*
*Boston, MA 02215*
Email: {*richwest,liye,missimer*}*@cs.bu.edu*

*Abstract*—Modern processors are increasingly featuring multiple cores, as well as support for hardware virtualization. While these processors are common in desktop and server-class computing, they are less prevalent in embedded and real-time systems. However, smartphones and tablet PCs are starting to feature multicore processors with hardware virtualization. If the trend continues, it is possible that future real-time systems will feature more sophisticated processor architectures. Future automotive or avionics systems, for example, could replace complex networks of uniprocessors with consolidated services on a smaller number of multicore processors. Likewise, virtualization could be used to isolate services and increase the availability of a system even when failures occur.

This paper investigates whether advances in modern processor technologies offer new opportunities to rethink the design of real-time operating systems. We describe some of the design principles behind Quest-V, which is being used as an exploratory vehicle for real-time system design on multicore processors with hardware virtualization capabilities. While not all embedded systems should assume such features, a case can be made that more robust, safety-critical systems can be built to use hardware virtualization without incurring significant overheads.

## I. INTRODUCTION

Multicore processors are becoming ubiquitous in all classes of computing, from desktops to servers, and even embedded systems. Many of these processors also support hardware virtualization capabilities. For example, Intel VT-x, AMD-V, and more recently, ARM Cortex A15 processors all have native machine virtualization support. While many such processors have been used in virtual datacenters, there is an opportunity to consider multicore virtualized systems in new areas of embedded computing. The ARM Cortex A15, for example, is being targeted at tablet devices and smartphones, with the ability to support multiple guest environments that separate personal and work-related information and services.

This paper investigates whether advances in modern processor technologies offer new opportunities to rethink the design of safety-critical real-time operating systems (RTOSes). We present a new system design that uses both virtualization capabilities and the redundancy offered by multiple processing cores, to develop a real-time system that is resilient to software faults. Our system, called Quest-V [1], is designed as a multikernel [2], or distributed system on a chip. It encapsulates different kernel instances, and their applications, in separate virtual machines (VMs). Faults in one VM are isolated from other VMs, similar to how processes are isolated from one another in a traditional operating system. However, each VM has greater capabilities than a traditional process running on top of a more privileged OS kernel. With Quest-V, each kernel instance in its own VM runs on top of a privileged monitor. While a monitor is required to be trusted, it can be kept to a minimal size and removed from the normal execution path of each kernel in the system.

Quest-V is not intended to replace RTOSes found in relatively simplistic *closed* embedded systems, with fixed tasksets and highly deterministic behavior. Instead, it is targeted at *open* real-time systems with dynamic tasksets and potentially unpredictable operating environments. Such systems often feature a mix of real-time and best-effort tasks, and data inputs generated by potentially untrusted external sources. Safety and security become significant, particularly in application domains such as health-care, factory automation, avionics and automotive systems. As an example, a future automotive system may not only involve internal tasks communicating over a controller area network, but may include dynamic tasks and data from interaction with other vehicles or the surroundings. External factors have significant consequences on the safe and timely operation of the vehicle [3].

While Linux supports hardware virtualization through its KVM interface, and hypervisors such as Xen [4] exist, neither approach has focused on providing real-time guarantees to tasks with safety and security constraints. Quest-V is an investigation into whether a single system can be built from a collection of separate kernel images operating together, to satisfy both temporal and spatial isolation requirements. Temporally, one task should not interfere with another in its timing requirements, while spatially, the misuse of resources such as memory by one task should not affect others.

We show how Quest-V does not incur significant operational overheads compared to a non-virtualized version of our system, simply called Quest, designed for SMP platforms. We describe how to enforce real-time guarantees on communication, interrupt handling, thread scheduling, and cross-core task migration. Similarly, we show how to leverage virtualization to prevent software component failures (either through error or malicious attacks) from compromising an entire system.

In the following section we describe the rationale for the design of Quest-V. This is followed by a description of the architecture in Section III. An experimental evaluation of the system is provided in Section IV. Here, we show the overheads of online fault recovery, along with the costs of using hardware virtualization to isolate kernels and system components. Section V describes related work, while conclusions are discussed in Section VI.

## II. DESIGN RATIONALE

Quest-V is centered around three main goals: safety, predictability and efficiency. The system is focused on safety-critical application domains, requiring high confidence in their operation [5] to prevent potential loss of lives or equipment. With recent advances in fields such as cyber-physical systems, more sophisticated OSes beyond those traditionally found in real-time and embedded computing are now required. Consider, for example, a collection of automotive sub-systems for engine, body, chassis, transmission, safety and infotainment services. These could be consolidated on the same multicore platform, with space-time partitioning to ensure malfunctions do not propagate across services. Moreover, if future automotive systems interact with their environments as part of vehicle-to-vehicle (V2V) or vehicle-to-infrastructure (V2I) communication, they are open to potential safety and security breaches from external sources.

While safety is a key goal, hardware virtualization provides a method to encapsulate, or *sandbox*, system resources from access by unauthorized sources. Virtualization provides an opportunity to enforce both system-wide safety and security beyond that achievable with non-virtualized hardware solutions such as paging and segmentation [6], [7]. Quest-V relies on *Extended Page Tables* (EPTs) [1] to separate system software components operating as a collection of services in a distributed system on a chip. The rationale for a virtualized multikernel is as follows:

*(1) Efficiency and Improved Predictability* – a multikernel adheres to the *share-nothing* principle, first discussed in the work on Barrelfish [2]. This leads to reduced resource contention and improved system efficiency on platforms with multiple cores, even accounting for explicit inter-kernel communication costs. As system resources are effectively distributed across cores, and each core is managed separately, there is no need to have shared structures such as a global scheduler queue. This, in turn, can improve predictability by eliminating undue blocking delays due to synchronization.

*(2) Fault Resilience* – replication of kernel functionality or, at least, separation of services in different protection domains increases fault resilience. This, in turn, increases system availability when there are partial system failures.

*(3) Highest Safe Privilege* – Rather than adopting a principle of *least* privilege for software services, as is done in micro-kernels, a virtualized system can support the *highest* safe privilege levels for different services. Virtualization provides an extra logical "ring of protection" that allows *guests* to think they are working directly on the hardware. Thus, virtualized services can be written with traditional kernel privileges, yet still be isolated from other equally privileged services in other guest domains. This avoids the costs typically associated with micro-kernels, which require added communication overheads to request services in different protection domains.

*(4) Minimal Trusted Code Base* – A micro-kernel attempts to provide a minimal trusted code base for the services it supports. However, it must still be accessed as part of inter-process communication, and basic operations such as coarse-grained memory management. Monitors form a trusted code base in the Quest-V virtualized multikernel. Access to these can be *avoided almost entirely*, except to bootstrap (guest) sandbox kernels, handle faults and manage EPTs. This enables sandboxes to operate, for the most part, independently of any other code base that requires trust. In turn, the trusted code base (i.e., monitors) can be limited to a small memory footprint.

While Quest-V uses hardware virtualization to isolate sandbox kernels, this is not a requirement of our multikernel approach. Many platforms, especially those in embedded systems, still lack hardware virtualization features. In such cases, it is possible to use alternative memory protection schemes based on segmentation or paging, for example. In this work, we seek to investigate the costs, and feasibility, of using hardware virtualization as a *first-class* feature of a chip-level distributed system. This contrasts with past work on virtual machines that treat guests as mostly separate and unrelated entities.

---

[1] Intel uses the term "EPT", while AMD refers to them as Nested Page Tables (NPTs). We use the term EPT for consistency.

## III. QUEST-V ARCHITECTURE

A high-level overview of the Quest-V architecture is shown in Figure 1. Each sandbox encapsulates a subset of machine physical resources (i.e., memory, one or more CPU cores and I/O devices), along with a kernel instance and its applications. A single hypervisor is replaced by a separate trusted monitor for each sandbox. This prevents a monitor from having to switch EPT mappings on return from handling VM-Exits [2], since the same guest, or sandbox kernel, will always resume. Additionally, separate monitors can be implemented differently, to prevent vulnerabilities to the same security threat.

Each monitor occupies less than a 4KB memory page. Apart from establishing EPT memory mappings [3] for sandboxes and communication channels, and assisting in fault recovery and migration, the monitors are not needed. Each sandbox kernel performs its own local scheduling and I/O handling without the cost of VM-Exits into a monitor. This is a significant departure from traditional virtual machine (VM) systems, which require the hypervisor to schedule guest VMs and manage I/O.

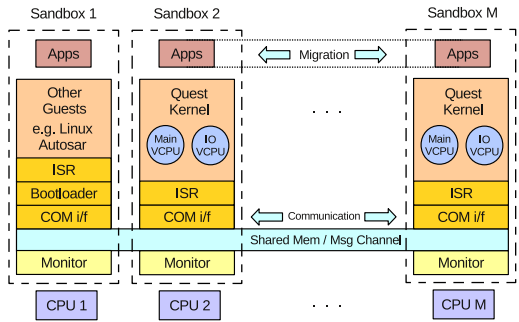

Fig. 1. Quest-V Architecture Overview

The extent to which functionality is separated across kernels is somewhat configurable in the Quest-V design. In our initial implementation, each sandbox kernel replicates most functionality, offering a private version of the corresponding services to its local application threads. It is, however, possible to have some kernels run Quest real-time services, while others run alternative kernels based on Linux or Autosar, for example.

Quest-V allows *any* sandbox to be configured for corresponding device interrupts, rather than having a dedicated sandbox for all communication with that device. This greatly reduces the communication and control paths necessary for I/O requests from applications in Quest-V. It also differs from the split-driver approach in systems such as Xen, which require all device interrupts to be channeled through a special driver domain. Sandboxes can be isolated from unnecessary drivers and services. Likewise, a sandbox can be provided with its own private set of devices and drivers, so if a software failure occurs in one driver, it will not affect all other sandboxes.

Quest-V allows each sandbox kernel to be configured to operate on a chosen subset of CPUs, or *cores*. This is similar to how Corey partitions resources amongst applications [8]. In our current approach, we assume each sandbox kernel is associated with one physical core since that simplifies local (sandbox) scheduling and allows for relatively easy enforcement of service guarantees using a variant of rate-monotonic scheduling [9]. Notwithstanding, application threads can be migrated between sandboxes as part of a load balancing strategy. Similarly, multi-threaded applications can be distributed across sandboxes to allow parallel thread execution.

Application and system services in distinct sandbox kernels can communicate via shared memory channels. Channels are established by EPT mappings setup by the corresponding monitors. Messages are passed across these channels similar to the approach in Barrelfish [2]. While a shared communication channel can be corrupted by a sandbox failure, EPTs prevent corruption of private memory regions in a remote sandbox. All other remote sandboxes with separate communication channels can continue to operate without compromise.

Main and I/O *virtual CPUs* (VCPUs) are used for real-time management of CPU cycles, to enforce *temporal isolation*. Application and system threads are bound to VCPUs, which in turn are assigned to underlying physical CPUs. We will discuss this further in the following section.

### A. System Implementation

Quest-V has been implemented from scratch as a 32-bit x86 system. Plans are underway to port the system to ARM Cortex A15 processors that also support hardware virtualization. The kernel code is approximately 10,000 lines of C and assembly, discounting drivers and network stack [10]. Using EPTs, each sandbox virtual address space is mapped to its own host memory region. By default, only the BIOS is shared across sandboxes, while all other functionality is privately mapped. EPT mappings can be established for shared communication channels between pairwise groups of sandboxes. Access from one sandbox into another sandbox's memory space is still, however, restricted to the pages of memory within this shared channel. Once bootstrapped, a Quest-V sandbox kernel can operate as a bootloader for a third-party system (e.g., a Linux or Autosar guest). This allows inter-operation between different sub-systems co-existing on the same hardware.

---
[2]E.g., due to an EPT violation caused by a fault.
[3]The page tables for EPTs take additional space but a 12KB mapping is enough for a 1GB sandbox.

**Hardware-Assisted Memory Isolation.** Figure 2 shows how address translation works for Quest-V guests (i.e., sandboxes) using Intel's extended page tables. Each sandbox kernel uses its own internal paging structures to translate guest virtual addresses to guest physical addresses (GPAs). EPT structures are then walked by the hardware to complete the translation to host physical addresses (HPAs). Modern processors with hardware support (e.g., Intel VT-x processors) avoid the need for software managed shadow page tables, and they also support TLBs to cache various intermediate translation stages. This greatly reduces the cost of address translation, as will be seen in Section IV-A.

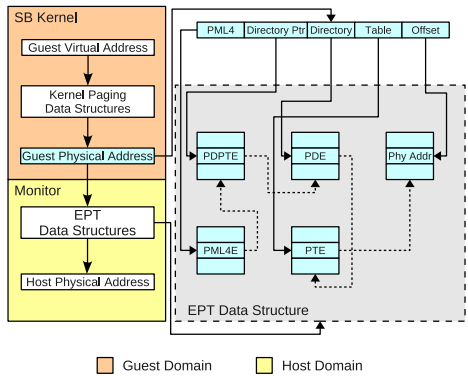

Fig. 2. Extended Page Table Mapping

On VT-x processors, address mappings can be manipulated at 4KB page granularity. For each 4KB page we have the ability to set read, write and even execute permissions. Consequently, attempts by one sandbox to access illegitimate memory regions of another will incur an EPT violation, causing a trap to the local monitor. The EPT data structures are, themselves, restricted to access by the monitors, thereby preventing tampering by sandbox kernels.

EPT support alone is actually insufficient to prevent faulty device drivers from corrupting the system. It is still possible for a malicious driver or a faulty device to DMA into arbitrary physical memory. This can be prevented with technologies such as Intel's VT-d, which restrict the regions into which DMAs can occur using IOMMUs. However, this is still insufficient to address other more insidious security vulnerabilities such as "white rabbit" attacks [11]. For example, a PCIe device can be configured to generate a Message Signaled Interrupt (MSI) with arbitrary vector and delivery mode by writing to *Local APIC memory*. Such malicious attacks can be addressed using hardware techniques such as Interrupt Remapping (IR), which restrict both the source and destination of interrupts.

**Real-Time VCPU Scheduling.** For use in real-time systems, the system must perform certain tasks by their deadlines. Quest-V does not require tasks to specify deadlines but instead ensures that the execution of one task does not interfere with the timely execution of others. For example, Quest-V is capable of scheduling interrupt handlers as threads, so they do not unduly interfere with the execution of higher-priority tasks. While Quest-V's scheduling framework is described elsewhere [12], we briefly explain how it provides temporal isolation between tasks and system events.

In Quest-V, VCPUs form the fundamental abstraction for scheduling and temporal isolation of the system. The concept of a VCPU is similar to that in virtual machines [13], [4], where a hypervisor provides the illusion of multiple *physical CPUs* (PCPUs) [4] represented as VCPUs to each of the guest virtual machines. VCPUs exist as kernel (*as opposed to monitor*) abstractions, to simplify the management of resource budgets for potentially many software threads. We use a hierarchical approach in which VCPUs are scheduled on PCPUs and threads are scheduled on VCPUs.

A VCPU acts as a resource container [14] for scheduling and accounting decisions on behalf of software threads. It serves no other purpose to virtualize the underlying physical CPUs, since our sandbox kernels and their applications execute directly on the hardware. In particular, a VCPU does not need to act as a container for cached instruction blocks that have been generated to emulate the effects of guest code, as in some trap-and-emulate virtualized systems.

In common with *bandwidth preserving* servers [15], [16], [17], each VCPU, $V$, has a maximum compute time budget, $C_V$, available in a time period, $T_V$. $V$ is constrained to use no more than the fraction $U_V = \frac{C_V}{T_V}$ of a physical processor (PCPU) in any window of real-time, $T_V$, while running at its normal (foreground) priority. To avoid situations where PCPUs are idle when there are threads awaiting service, a VCPU that has expired its budget may operate at a lower (background) priority. All background priorities are set below those of foreground priorities to ensure VCPUs with expired budgets do not adversely affect those with available budgets.

[4]We define a PCPU to be either a conventional CPU, a processing core, or a hardware thread.

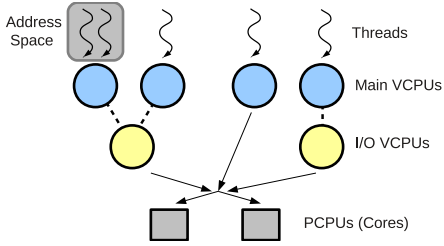

Fig. 3. VCPU Scheduling Hierarchy

Quest-V defines two classes of VCPUs as shown in Figure 3: (1) *Main VCPUs* are used to schedule and track the PCPU usage of conventional software threads, while (2) *I/O VCPUs* are used to account for, and schedule the execution of, interrupt handlers for I/O devices. This distinction allows for interrupts from I/O devices to be scheduled as threads, which may be deferred execution when threads associated with higher priority VCPUs having available budgets are runnable. The flexibility of Quest-V allows I/O VCPUs to be specified for certain devices, or for certain tasks that issue I/O requests, thereby allowing interrupts to be handled at different priorities and with different CPU shares than conventional tasks associated with Main VCPUs.

By default, each Main VCPU acts like a Sporadic Server [18], [19], while each I/O VCPU acts as a bandwidth preserving server with a dynamically-calculated period, $T_{IO}$, and budget, $C_{IO}$ [12]. Each I/O VCPU is specified a certain utilization factor, $U_{IO}$, to limit its bandwidth. When a device interrupt requires handling by an I/O VCPU, the system determines the thread $\tau$ associated with a corresponding I/O request [5]. In Quest-V, all events including those related to I/O processing are associated with threads running on Main VCPUs. $C_{IO}$ is calculated as $T_V \cdot U_{IO}$, while $T_{IO}$ is set to $T_V$ for a Main VCPU, $V$, associated with $\tau$.

Figure 4 shows an example schedule for two Main VCPUs and one I/O VCPU, with a utilization factor of 4%, for a certain device such as a gigabit Ethernet card. Replenishment lists are shown for VCPU1. Since the I/O VCPU handles I/O requests on behalf of a thread running on VCPU1, it inherits a budget, $C_{IO} = 50*0.04$, and period, $T_{IO} = 50$.

The invariant is that the sum of replenishment amounts for all list items must not exceed the budget capacity of the corresponding VCPU (here, 20, for VCPU1). Also, no future replenishment, $R$, for a VCPU, $V$, executing from $t$ to $t+R$ can occur before $t+T_V$ [19].

[5]E.g., $\tau$ may have issued a prior `read()` request that caused it to block on its Main VCPU, but which ultimately led to a device performing an I/O operation.

**Temporal Isolation.** In Quest-V, VCPUs are mapped to a separate scheduling queue for each PCPU. Under this arrangement, our default policies for Main and I/O VCPU scheduling allow us to guarantee temporal isolation if the Liu-Layland utilization bound is satisfied [9]. For a single PCPU with $n$ Main VCPUs and $m$ I/O VCPUs we have the following:

$$\sum_{i=0}^{n-1} \frac{C_i}{T_i} + \sum_{j=0}^{m-1} (2-U_j) \cdot U_j \leq n\left(\sqrt[n]{2}-1\right) \quad (1)$$

Here, $C_i$ and $T_i$ are the budget capacity and period of Main VCPU $V_i$, and $U_j$ is the utilization factor of I/O VCPU $V_j$. Further details are available outside this paper [12]. This bound can be improved with dynamic priority scheduling of VCPUs (e.g., using earliest deadline first scheduling) but this adds more overhead to the scheduler. This is because: (1) dynamic priorities require more complex queue management, and (2) Quest-V uses Local APIC timers, programmed for one-shot operation, to trigger an interrupt in time for the next event to be processed; more frequent reprogramming of timers may be necessary if priorities change.

Quest-V admission control uses Equation 1 to decide whether to allow the creation of a new VCPU. In overload conditions, static priority scheduling has the advantage that the highest priority subset of VCPUs capable of meeting their timing requirements will not be affected by lower priority VCPUs. This is not the case with dynamic priority scheduling, where overload can cause all VCPUs to fail to maintain their correct PCPU shares. Similarly, hypervisor scheduling using policies such as Borrowed Virtual Time (BVT) [20] cannot guarantee temporal isolation between VCPUs over specific real-time windows.

**Real-Time Communication.** Inter-sandbox communication in Quest-V relies on message passing primitives built on shared memory, and asynchronous event notification mechanisms using Inter-processor Interrupts (IPIs). IPIs are currently used to communicate with remote sandboxes to assist in fault recovery, and can also be used to notify the arrival of messages exchanged via shared memory channels. Monitors update extended page table mappings as necessary to establish message passing channels between specific sandboxes. Only those sandboxes with mapped shared pages are able to communicate with one another.

A *mailbox* data structure is set up within shared memory by each end of a communication channel. By default, Quest-V currently supports asynchronous communication by polling a status bit in each relevant mailbox to determine message arrival. Message passing threads are bound to VCPUs with specific parameters to control the rate of exchange of information. Likewise, sending and receiving threads are assigned to higher priority VCPUs to reduce the latency of transfer of information across a communication channel. This way, shared memory channels can be prioritized and granted higher or lower throughput as needed, while ensuring information is communicated in a predictable manner. Thus, Quest-V supports real-time communication between sandboxes without compromising the CPU shares allocated to non-communicating tasks.

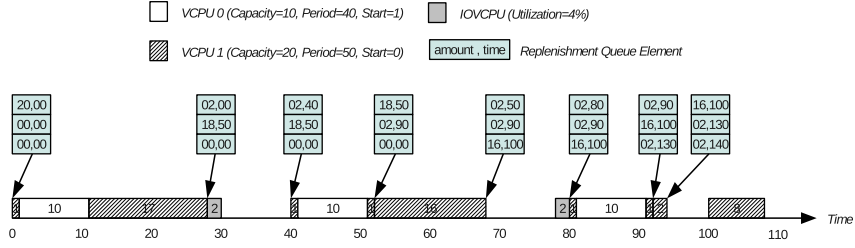

Fig. 4. Example VCPU Schedule

**Predictable Migration.** Quest-V restricts migratable address spaces to those associated with VCPUs that either: (1) have currently expired budgets, or (2) are waiting in a sleep queue. In the former case, the VCPU is not runnable at its foreground priority until its next budget replenishment. In the latter case, a VCPU is blocked until a wakeup event occurs (e.g., due to an I/O request completion or a resource becoming available). Together, these two cases prevent migrating a VCPU when it is runnable, as the migration delay could impact the VCPU's utilization.

For VCPU, $V_s$, associated with a migrating address space, we define $E_s$ to be the *relative time* [6] of the next event, which is either a replenishment or wakeup. For the utilization of $V_s$ to be unaffected by migration, the following must hold:

$$E_s \geq \lfloor \frac{\Delta_s}{C_m} \rfloor \cdot T_m + \Delta_s \bmod C_m, \quad (2)$$

where $C_m$ and $T_m$ are the budget and period of the migrating thread's VCPU, and $\Delta_s$ is the migration cost of copying an address space and its *quest_tss* data structures to the destination. At boot time, Quest-V establishes base costs for copying memory pages without caches enabled [7]. These costs are used to determine $\Delta_s$ for a given address space size. Quest-V makes sure that the migrating thread will not be woken up by asynchronous events until the migration is finished. The system imposes the restriction that threads waiting on I/O events cannot be migrated.

A schedulability test is performed before migration, to ensure a VCPU can be added to the destination sandbox without affecting total utilization. If the test fails, the migration request will be rejected immediately by an IPI. A VCPU can be migrated immediately for any successful test, if it does not require its utilization to be guaranteed while migration is in progress.

In order to simplify the migration criteria, our current implementation restricts concurrent migration requests to different destination sandboxes. This is not problematic as migrations are expected to be infrequent.

**Clock Synchronization.** One extra challenge to be considered during migration is clock synchronization between different sandboxes. Quest-V schedulers use Local APIC Timers and Time Stamp Counters (TSCs) in each core as the source for all time-related activities in the system, and these are not guaranteed to be synchronized by hardware. Consequently, Quest-V adjusts time for each migrating address space to compensate for clock skew. This is necessary when updating budget replenishment and wakeup time events for a migrating VCPU that is sleeping on an I/O request, or which is not yet runnable.

The source sandbox places its current TSC value in shared memory immediately before sending an IPI migration request. This value is compared with the destination TSC when the IPI is received. A time-adjustment, $\delta_{ADJ}$, for the migrating VCPU is calculated as follows:

$$\delta_{ADJ} = TSC_d - TSC_s - 2 * RDTSC_{cost} - IPI_{cost} \quad (3)$$

$TSC_d$ and $TSC_s$ are the destination and source TSCs, while $RDTSC_{cost}$ and $IPI_{cost}$ are the average costs of reading a TSC and sending an IPI, respectively. $\delta_{ADJ}$ is then added to all future budget replenishment and wakeup time events for the migrating VCPU in the destination sandbox.

**Interrupt Distribution and I/O Management.** By default, Quest-V allows interrupts to be delivered di-

---
[6] i.e., Relative to current time.
[7] We do not consider memory bus contention issues, which could make worst-case estimations even larger.

rectly to sandbox kernels. Hardware interrupts are delivered to *all* sandbox kernels with access to the corresponding device. This avoids the need for interrupt handling to be performed in the context of a monitor as is typically done with conventional virtual machine approaches. Experiments show that virtualization does not add significant overheads for handling interrupts or I/O requests. See Section VII-C in the Appendix for further details.

## IV. EXPERIMENTAL EVALUATION

We conducted a series of experiments to investigate the performance, predictability and fault isolation of Quest-V. For network experiments, we ran Quest-V on a mini-ITX machine with a Core i5-2500K 4-core processor, featuring 8GB RAM and a Realtek 8111e NIC. In all other cases we used a Dell PowerEdge T410 server with an Intel Xeon E5506 2.13GHz 4-core processor, featuring 4GB RAM. Unless otherwise stated, all software threads were bound to Main VCPUs with 100% total utilization for performance related experiments.

### A. Address Translation Overhead

To show the costs of address translation as described in Figure 2, we measured the latency to access a number of data and instruction pages in a guest user-space process. Figures 5 and 6 show the execution time of a process bound to a Main VCPU with a 20ms budget every 100ms. Instruction and data references to consecutive pages are 4160 bytes apart to avoid cache aliasing effects. The results show the average cost to access working sets taken over 10 million iterations. In the cases where there is a TLB flush or a VM exit, these are performed each time the set of pages on the x-axis has been referenced.

For working sets less than 512 pages Quest-V (`Base` case) performs as well as a non-virtualized version of Quest. Extra levels of address translation with extended paging only incur costs above the two-level paging of a 32-bit Quest virtual memory system when address spaces are larger than 512 pages. For embedded systems, we do not see this as a limitation, as most applications will have smaller working sets. As can be seen, the costs of a VM-Exit are equivalent to a TLB flush, but Quest-V avoids this by operating more in common with the `Quest-V Base` case. Hence, extended paging does not incur significant overheads under normal circumstances, as the hardware TLBs are being used effectively.

### B. Fault Isolation and Predictability

To demonstrate fault isolation in Quest-V, we created a scenario that includes both message passing and network service across 4 different sandboxes. Specifically, sandbox 1 has a kernel thread that sends messages through private message passing channels to sandbox 0, 2 and 3. Each private channel is shared only between the sender and specific receiver, and is guarded by EPTs. In addition, sandbox 0 also has a network service running that handles ICMP echo requests. After all the services are up and running, we manually break the NIC driver in sandbox 0, overwrite sandbox 0's message passing channel shared with sandbox 1, and try to corrupt the kernel memory of other sandboxes to simulate a driver fault. After the driver fault, sandbox 0 will try to recover the NIC driver along with both network and message passing services running in it. During the recovery, the whole system activity is plotted in terms of message reception rate and ICMP echo reply rate in all available sandboxes and the results are shown in Figure 7.

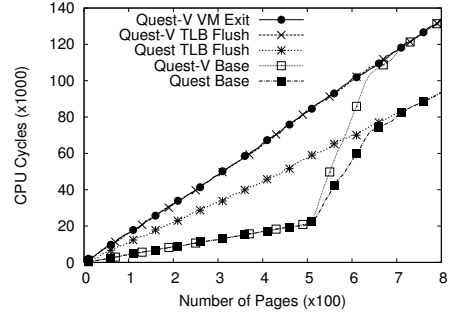

Fig. 5. Data TLB Performance

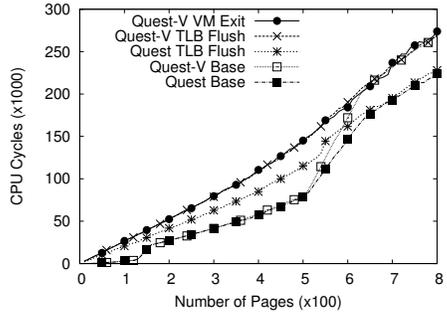

Fig. 6. Instruction TLB Performance

In the experiment, sandbox 1 broadcasts messages to others at 50 millisecond intervals, while sandbox 0, 2 and 3 receive at 100, 800 and 1000 millisecond intervals. Also, another machine in the local network sends ICMP echo requests at 500 millisecond intervals to sandbox 0. All message passing threads are bound to Main VCPUs with 100ms periods and 20% utilization. The network driver thread is bound to an I/O VCPU with 10% utilization and 10ms period.

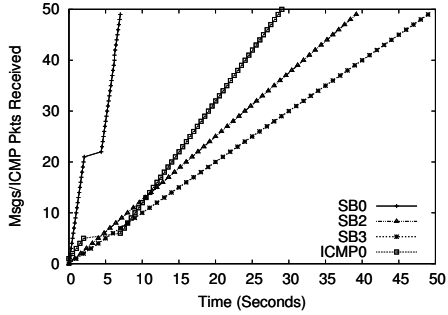

Fig. 7. Sandbox Isolation

Results show that an interruption of service happened for both message passing and network packet processing in sandbox 0, but all the other sandboxes were unaffected. This is because of memory isolation between sandboxes enforced by EPTs.

Finally, we ran an experiment using the TORCS driving simulator [21]. The AI engine for a self-driving car was first run in a Quest-V sandbox and then in an Ubuntu Linux 10.04 guest on Xen 4.1.3. In each case, we simulated a fault that either affected an entire Quest-V sandbox or Linux kernel. A "hot standby" version of the AI engine also ran in a separate sandbox or, for Linux, a different guest. Fault recovery triggered the hot standby, which ran in a domain with a CPU hog attempting to consume as much time as possible. For Linux, the hog prevented the hot standby running as desired. For Quest-V, the hog was restricted to a VCPU that allowed the AI engine to acquire 2ms every 5ms of CPU time. Consequently, Figure 8 shows the car was able to drive smoothly around the track without the effects of faulting services or competition for CPU resources in Quest-V. For Linux on Xen, the vehicle's trajectory was affected as shown in the right-most figure. Other scenarios, not shown, led to the Linux AI engine suffering enough delay to cause the vehicle to crash.

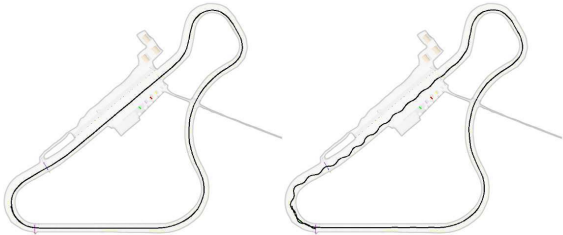

Fig. 8. TORCS: Quest-V (left) and Linux-Xen (right)

The lack of real-time predictability in Linux means there is no temporal isolation between a CPU hogging thread and the AI engine. Similarly, the underlying hypervisor is also not able to guarantee real-time CPU shares to its Linux guests. These factors cause the vehicle's trajectory to be affected when a system fault occurs. Moreover, the Linux guests are not running on a hypervisor that is aware they are cooperating in fault recovery. Consequently, the TORCS AI engine and the hot standbys must be written in a way to be aware of each others existence. Minimally, this requires filtering of the hot standby's messages exchanged with the server when the primary AI engine is operational. In contrast, the Quest-V driver layer can be configured to filter data from hot standbys until they are needed in fault recovery. This means applications do not have to be written explicitly to coordinate with fault recovery processes.

### C. Predictable Migration and Communication

To verify the predictability of the Quest-V migration framework, we constructed a task group consisting of 2 communicating threads and another CPU-intensive thread running a Canny edge detection algorithm on a stream of video frames. The frames were gathered from a LogiTech QuickCam Pro9000 camera mounted on our RacerX mobile robot, which traversed one lap of Boston University's indoor running track at Agganis Arena [8]. To avoid variable bit rate frames affecting the results of our experiments, we applied Canny repeatedly to the frame shown in Figure 9 rather than a live stream of the track. This way, we could determine the effects of migration on a Canny thread by observing changes in processing rate while the other threads communicated with each other.

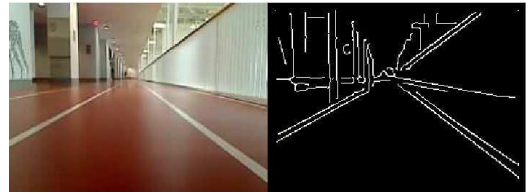

Fig. 9. Track Image Processed by Canny

For all the experiments in this section, we have two active sandbox kernels each with 5 VCPUs. The setup is shown in Table I. The Canny thread is the target for migration from sandbox 1 to sandbox 2 in all cases. Migration always starts at time 5. A logger thread was used to collect the result of the experiment in a predictable manner. Data points are sampled and reported in a one second interval.

Figure 10 shows the behavior of Canny as it is migrated in the presence of the two communicating threads. The left y-axis shows both Canny frame rate (in frames-per-second, *fps*) and message passing throughput (in multiples of a 1000 Kilobytes-per-second). The right y-axis shows the actual CPU consumption of the migration thread in (millions of, *x1m*) cycles. We can see from this figure that none of the threads (2 communicating threads and Canny) have been affected due to migration. The sudden spike in migration thread CPU consumption occurs during the migration of the Canny thread.

---

[8]RacerX is a real-time robot control project that leverages Quest-V.

| VCPU (C/T) | Sandbox 1 | Sandbox 2 |
|---|---|---|
| 20/100 | Shell | Shell |
| 10/50 | Migration Thread | Migration Thread |
| 20/100 | Canny | |
| 20/100 | Logger | Logger |
| 10/100 | Comms 1 | Comms 2 |

TABLE I
MIGRATION EXPERIMENT VCPU SETUP

The average time to migrate an address space varies from under 1 millisecond to about 10 - 20 milliseconds, depending on the actual address space size. This is with all caches enabled and with address spaces being limited to a maximum of $4MB$.

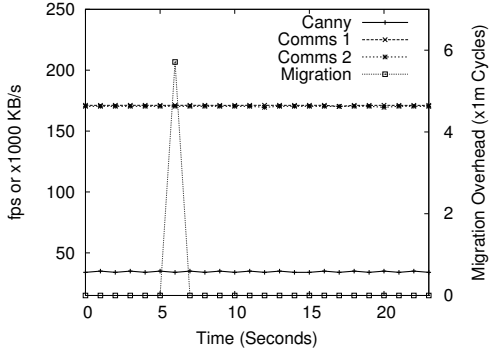

Fig. 10. Migration With No Added Overhead

Table II shows the values of variables as defined in Equation 2. The worst-case migration cost, $\Delta_s, worst$, was the cost of copying a Canny address space with all caches disabled (including the overhead of walking its page directory). $\Delta_s, actual$ was the actual migration thread budget consumption during migration. Both worst-case and actual migration costs satisfy the constraints of Equation 2, while considering Equation 3. Consequently, all VCPUs remain unaffected in terms of their CPU utilization during migration.

| Variables | $E_s$ | $\Delta_s, worst$ | $\Delta_s, actual$ | $C_m$ | $T_m$ |
|---|---|---|---|---|---|
| Time (ms) | 79.8 | 5.4 | 1.7 | 10 | 50 |

TABLE II
MIGRATION CONDITION

## V. RELATED WORK

Barrelfish[2] is a multikernel that replicates system state rather than sharing it, to avoid the costs of synchronization and management of shared data structures. As with Quest-V, communication between kernels is via explicit message passing, using shared memory channels to transfer cache-line-sized messages. In contrast to Barrelfish, Quest-V uses virtualization mechanisms to isolate kernel services as part of our goal to develop high-confidence systems.

Factored OS (FOS) [22] and Corey [8] are partitioning systems that distribute OS services and applications across processing cores. These systems are focused on scalability rather than Quest-V's primary focus on fault isolation and predictability.

There have been several systems that rely on virtualization techniques to enforce logical isolation and implement scalable resource management on multicore and multiprocessor platforms. Cellular Disco [23] extends the Disco virtual machine monitor (VMM) [24] with support for hardware fault containment. As with Hive [25], the system is partitioned into cells, each containing a copy of the monitor code and all machine memory pages belonging to the cell's nodes. Failure of one cell only affects VMs using resources in that cell.

Xen[4] is a subsequent VMM that uses a special driver domain and (now optional) para-virtualization techniques to support multiple guests. In contrast to VMMs such as Disco, Xen and also the Wind River Hypervisor [26], Quest-V operates as a single system with sandbox kernels potentially implementing different services that are isolated using memory virtualization. Quest-V also avoids the need for a split-driver model involving a special domain (e.g., `Dom0` in Xen) to handle device interrupts.

Finally, PikeOS [27] is a separation micro-kernel [28] that supports multiple guest VMs, and targets safety-critical domains such as Integrated Modular Avionics. The micro-kernel supports a virtualization layer that is required to manage the spatial and temporal partitioning of resources amongst guests. This contrasts with Quest-V, where each sandbox kernel is responsible for its own local scheduling on a subset of processor cores – a VMM, or hypervisor, is *not* required for scheduling.

## VI. CONCLUSIONS

This paper describes the Quest-V multikernel, designed for real-time, safety-critical systems. Extended page tables are used to isolate sandbox kernels across different cores in a multicore system. Hardware virtualization provides a fault containment mechanism via an extra logical "ring of protection". This enables untrusted software to run with traditional kernel-level privileges, without compromising the entire system. This contrasts with the micro-kernel approach of providing the least privileges necessary to software, and using the kernel as a trusted communication channel for interaction between less trusted software components. Quest-V, for the most part, eliminates the trusted component (here, a monitor) from the critical path of software execution. Monitors are only needed for fault handling, migration, and updating EPT mappings (e.g., to establish communication channels with other sandboxes).

Experiments show that hardware virtualization does not add significant overheads in our design, as VM-Exits into monitor code are not normally needed. Unlike conventional hypervisors that virtualize underlying hardware for use by multiple disparate guests, Quest-V assumes all sandboxes are operating together as one collective distributed system on a chip. Each sandbox kernel is responsible for scheduling of its threads and VCPUs onto local hardware cores. Local scheduling within each sandbox involves the management of time budgeted, temporally-isolated VCPUs. Similarly, memory allocation and I/O management are handled within each sandbox *without* involvement of a monitor. While monitors must be trusted, they are rarely accessed and have a small code base. Moreover, using technologies such as Intel's Trusted Execution Technology (TXT), it is possible to enforce safety of the monitors themselves.

See http://questos.github.com for more information.

REFERENCES

[1] R. West, Y. Li, and E. Missimer, "Time management in the Quest-V RTOS," in *Proceedings of the 8th Annual Workshop on Operating Systems Platforms for Embedded Real-Time Applications (OSPERT)*, (Pisa, Italy), July 2012.
[2] A. Baumann, P. Barham, P.-E. Dagand, T. Harris, R. Isaacs, S. Peter, T. Roscoe, A. Schüpbach, and A. Singhania, "The Multi-kernel: A new OS architecture for scalable multicore systems," in *Proceedings of the 22nd ACM Symposium on Operating Systems Principles*, pp. 29–44, 2009.
[3] S. Checkoway, D. McCoy, B. Kantor, D. Anderson, H. Shacham, S. Savage, K. Koscher, A. Czeskis, F. Roesner, and T. Kohno, "Comprehensive experimental analyses of automotive attack surfaces," in *Proceedings of the 20th USENIX Conference on Security*, 2011.
[4] P. Barham, B. Dragovic, K. Fraser, S. Hand, T. Harris, A. Ho, R. Neugebauer, I. Pratt, and A. Warfield, "Xen and the art of virtualization," in *Proceedings of the 19th ACM Symposium on Operating Systems Principles*, pp. 164–177, 2003.
[5] "NITRD Working Group: IT Frontiers for a New Millenium: High Confidence Systems," April 1999.
[6] T. Chiueh, G. Venkitachalam, and P. Pradhan, "Integrating segmentation and paging protection for safe, efficient and transparent software extensions," in *Proceedings of the 17th ACM Symposium on Operating Systems Principles*, pp. 140–153, 1999.
[7] V. Uhlig, U. Dannowski, E. Skoglund, A. Haeberlen, and G. Heiser, "Performance of address-space multiplexing on the Pentium," Tech. Rep. 2002-1, University of Karlsruhe, Germany, 2002.
[8] S. Boyd-Wickizer, H. Chen, R. Chen, Y. Mao, M. F. Kaashoek, R. Morris, A. Pesterev, L. Stein, M. Wu, Y. hua Dai, Y. Zhang, and Z. Zhang, "Corey: An operating system for many cores," in *Proceedings of the 8th USENIX Symposium on Operating Systems Design and Implementation*, pp. 43–57, 2008.
[9] C. L. Liu and J. W. Layland, "Scheduling algorithms for multiprogramming in a hard-real-time environment," *Journal of the ACM*, vol. 20, no. 1, pp. 46–61, 1973.
[10] "Quest." http://www.cs.bu.edu/fac/richwest/quest.html.
[11] R. Wojtczuk and J. Rutkowska, "Following the white rabbit: Software attacks against Intel VT-d technology," April 2011. Inivisible Things Lab.
[12] M. Danish, Y. Li, and R. West, "Virtual-CPU Scheduling in the Quest Operating System," in *Proceedings of the 17th Real-Time and Embedded Technology and Applications Symposium*, pp. 169–179, 2011.
[13] K. Adams and O. Agesen, "A comparison of software and hardware techniques for x86 virtualization," in *Proceedings of the 12th Intl. Conf. on Architectural Support for Programming Languages and Operating Systems*, pp. 2–13, 2006.
[14] G. Banga, P. Druschel, and J. C. Mogul, "Resource Containers: A new facility for resource management in server systems," in *Proceedings of the 3rd USENIX Symposium on Operating Systems Design and Implementation*, 1999.
[15] L. Abeni, G. Buttazzo, S. Superiore, and S. Anna, "Integrating multimedia applications in hard real-time systems," in *Proceedings of the 19th IEEE Real-time Systems Symposium*, 1998.
[16] Z. Deng, J. W. S. Liu, and J. Sun, "A scheme for scheduling hard real-time applications in open system environment," in *the 9th Euromicro Workshop on Real-Time Systems*, 1997.
[17] M. Spuri, G. Buttazzo, and S. S. S. Anna, "Scheduling aperiodic tasks in dynamic priority systems," *Real-Time Systems*, vol. 10, pp. 179–210, 1996.
[18] B. Sprunt, L. Sha, and J. Lehoczky, "Aperiodic task scheduling for hard real-time systems," *Real-Time Systems Journal*, vol. 1, no. 1, pp. 27–60, 1989.
[19] M. Stanovich, T. P. Baker, A.-I. Wang, and M. G. Harbour, "Defects of the POSIX sporadic server and how to correct them," in *Proceedings of the 16th IEEE Real-Time and Embedded Technology and Applications Symposium*, 2010.
[20] K. J. Duda and D. R. Cheriton, "Borrowed-virtual-time (bvt) scheduling: supporting latency-sensitive threads in a general-purpose scheduler," in *Proceedings of the 7th ACM Symposium on Operating Systems Principles*, pp. 261–276, 1999.
[21] "TORCS." http://torcs.sourceforge.net/.
[22] D. Wentzlaff and A. Agarwal, "Factored operating systems (FOS): The case for a scalable operating system for multicores," *SIGOPS Operating Systems Review*, vol. 43, pp. 76–85, 2009.
[23] K. Govil, D. Teodosiu, Y. Huang, and M. Rosenblum, "Cellular Disco: Resource management using virtual clusters on shared-memory multiprocessors," in *Proceedings of the 17th ACM Symposium on Operating Systems Principles*, pp. 154–169, 1999.
[24] E. Bugnion, S. Devine, and M. Rosenblum, "Disco: Running commodity operating systems on scalable multiprocessors," in *Proceedings of the 16th ACM Symposium on Operating Systems Principles*, pp. 143–156, 1997.
[25] J. Chapin, M. Rosenblum, S. Devine, T. Lahiri, D. Teodosiu, and A. Gupta, "Hive: Fault containment for shared-memory multiprocessors," in *Proceedings of the 15th ACM Symposium on Operating Systems Principles*, pp. 12–25, 1995.
[26] http://www.windriver.com/products/hypervisor/.
[27] "Safe and secure virtualization in a separation microkernel." SYSGO AG, White Paper (www.sysgo.com).
[28] G. Klein, K. Elphinstone, G. Heiser, J. Andronick, D. Cock, P. Derrin, D. Elkaduwe, K. Engelhardt, R. Kolanski, M. Norrish, T. Sewell, H. Tuch, and S. Winwood, "seL4: Formal verification of an OS kernel," in *Proceedings of the 22nd ACM Symposium on Operating Systems Principles*, pp. 207–220, 2009.
[29] "lwIP." http://savannah.nongnu.org/projects/lwip/.

## VII. APPENDIX

Quest-V is designed to be robust against software faults that could potentially compromise a system kernel. As long as the integrity of one sandbox is maintained it is theoretically possible to build a Quest-V multikernel capable of recovering service functionality online. This contrasts with a traditional system approach, which may require a full system reboot if the kernel is compromised by faulty software such as a device driver.

Although fault detection mechanisms are not necessarily straightforward, faults are easily detected in Quest-V if they generate EPT violations. EPT violations transfer control to a corresponding monitor where they may be handled. More elaborate schemes for identifying faults will be covered in our future work.

Quest-V allows for fault recovery either in the local sandbox, where the fault occurred, or in a remote sandbox that is presumably unaffected. Upon detection of a fault, a method for passing control to the local monitor is required. If the fault does not automatically trigger a VM-Exit, it can be forced by a fault handler issuing an appropriate instruction. [9]

### A. Fault Recovery

To demonstrate the cost of fault recovery in Quest-V, we intentionally corrupted the NIC driver on the mini-ITX machine while running a simple HTTP 1.0-compliant web server in user-space. Our web server was ported to a socket API that we implemented on top of lwIP [29]. A remote Linux machine running `httperf` attempted to send 120 requests per second during both the period of driver failure and normal web server operation. Request URLs referred to the Quest-V website, with a size of 17675 bytes.

Figure 11 shows the request and response rate at 0.5s sampling intervals. The driver failure occurred in the interval [1.5s,2s], after which recovery took place. Recovery involved re-initializing the NIC driver and restarting the web server in another sandbox, taking less than 0.5s. This is significantly faster than a system reboot, which can take tens of seconds (or longer) to restart the network service.

Fault recovery can occur locally or remotely. In this experiment, we saw little difference in the cost of either approach. Either way, the NIC driver needs to be re-initialized. This either involves re-initialization of the same driver that faulted in the first place, or an alternative driver that is tried and tested. As fault detection is not in the scope of this paper, we triggered the fault recovery event manually by assuming an error occurred. Aside from optional replacement of the faulting driver, and re-initialization, the network interface needs to be restarted. This involves re-registering the driver with lwIP and assigning the interface an IP address. Table III shows the time for different phases of kernel-level recovery.

[9]For example, on the x86, the `cpuid` instruction forces a VM-Exit.

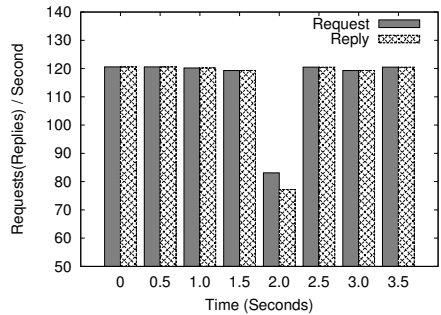

Fig. 11. Web Server Recovery

| Phases | CPU Cycles | |
|---|---|---|
| | Local Recovery | Remote Recovery |
| VM-Exit | 885 | |
| Driver Replacement | 10503 | N/A |
| IPI Round Trip | N/A | 4542 |
| VM-Enter | 663 | |
| Driver Re-initialization | 1.45E+07 | |
| Network I/F Restart | 78351 | |

TABLE III
OVERHEAD OF DIFFERENT PHASES IN FAULT RECOVERY

### B. Forkwait Microbenchmark

We measured the overhead of making system calls within Quest-V sandboxes to identify any costs associated with virtualization. Using a version of the `forkwait` microbenchmark [13], 40000 new processes were forked in each set of experiments, and the total CPU cycles were recorded. We then compared the performance of Quest-V against a version of Quest without hardware virtualization enabled, as well as a Linux 2.6.32 kernel in both 32- and 64-bit configurations. Results in Figure 12 suggest that hardware virtualization does not add any obvious overhead to Quest-V system calls. Moreover, both Quest and Quest-V took less time than Linux to complete their executions.

### C. Interrupt and I/O Management

Besides system calls, device interrupts also require control to be passed to a kernel. We therefore conducted a series of experiments to show the overheads of interrupt delivery and handling in Quest-V. For comparison, we recorded the number of interrupts that occurred and the *total* round-trip time to process 30000 ping packets on both Quest and Quest-V machines. A single (sandbox) kernel handled all interrupts in each case. Additionally, all ICMP requests were issued in 3 millisecond intervals from a remote machine. Results in Table IV show that virtualization does not affect performance.

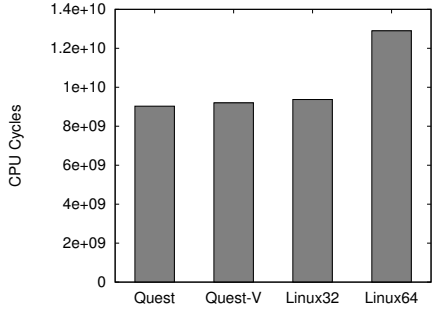

Fig. 12. Forkwait Microbenchmark

|  | Quest | Quest-V |
| --- | --- | --- |
| # Interrupts | 30004 | 30003 |
| Total round-trip time (ms) | 5737 | 5742 |

TABLE IV
INTERRUPT DISTRIBUTION AND HANDLING OVERHEAD

Figure 13 shows UDP throughput measurements using `netperf`, which was ported to the Quest-V and non-virtualized `Quest-SMP` systems. Up to 4 netperf clients were run in separate guest domains, or sandboxes, for all virtualized scenarios. We compared against para- (PVM) and hardware-virtualized (HVM) Xen 4.1.2 supporting Ubuntu 10.04 guests, as well as a non-virtualized Ubuntu 10.04 Linux system, with 1-4 netperf processes. In Xen and Linux the netperf clients were free to run on any core of the (Core i5) processor. Each client produced a stream of 16KB messages. Although inferior to non-virtualized Linux, Quest-V throughput is close to that of Quest-SMP and better than the admittedly non-optimized Xen system. Improvements to the network stack and Ethernet driver will hopefully bring Quest-V performance closer to Linux.

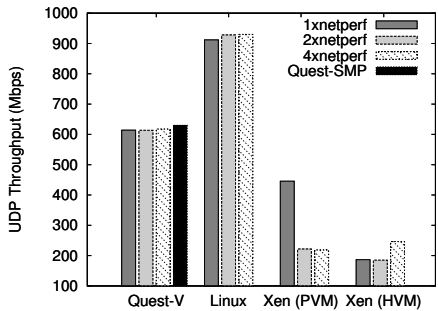

Fig. 13. UDP Throughput

### D. Inter-Sandbox Communication

The message passing mechanism in Quest-V is built on shared memory. Instead of focusing on memory and cache optimization, we tried to study the impact of scheduling on inter-sandbox communication in Quest-V.

We setup two kernel threads in two different sandbox kernels and assigned a VCPU to each of them. One kernel thread used a 4KB shared memory message passing channel to communicate with the other thread. In the first case, the two VCPUs were the highest priority with their respective sandbox kernels. In the second case, the two VCPUs were assigned lower utilizations and priorities, to identify the effects of VCPU parameters (and scheduling) on the message sending and receiving rates. In both cases, the time to transfer messages of various sizes across the communication channel was measured. Note that the VCPU scheduling framework ensures that all threads are guaranteed service as long as the total utilization of all VCPUs is bounded according to rate-monotonic theory [9]. Consequently, the impacts of message passing on overall system predictability can be controlled and isolated from the execution of other threads in the system.

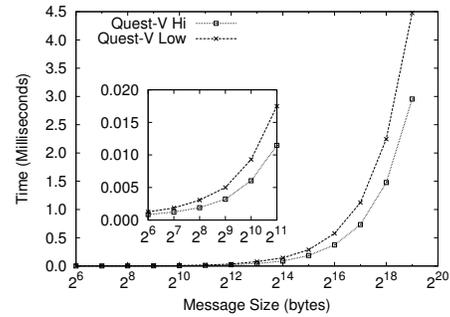

Fig. 14. Message Passing Microbenchmark

Figure 14 shows the time spent exchanging messages of various sizes, plotted on a log scale. *Quest-V Hi* is the plot for message exchanges involving high-priority VCPUs having 100ms periods and 50% utilizations for both the sender and receiver. *Quest-V Low* is the plot for message exchanges involving low-priority VCPUs having 100ms periods and 40% utilizations for both the sender and receiver. In the latter case, a shell process was bound to a highest priority VCPU. As can be seen, the VCPU parameters affect message transfer times.

In our experiments, the time spent for each size of message was averaged over a minimum of 5000 trials to normalize the scheduling overhead. The communication costs grow linearly with increasing message size, because they include the time to access memory.